\documentclass[11pt,a4paper]{article}
\usepackage{amsfonts}
\usepackage{amssymb}
\usepackage{amssymb,latexsym,amsmath,amsfonts,amsthm}
\usepackage{graphicx}
\usepackage{tikz}
\usepackage{pgflibraryshapes}
\usetikzlibrary{arrows,decorations.markings}
\usepackage{epsfig}
\usepackage{comment,verbatim}
\usepackage{hyperref}

\newcommand{\ud}{\,\mathrm{d}}

\def\supp{\mathop{\mathrm{supp}}\nolimits}

\newtheorem{theorem}{Theorem}[section]

\newtheorem{proposition}[theorem]{Proposition}

\theoremstyle{definition}

\theoremstyle{remark}

\newtheorem{remark}[theorem]{Remark}

\numberwithin{equation}{section}

\begin{document}
\title{A note on the limiting mean distribution of
singular values for products of two Wishart random matrices}

\author{Lun Zhang\footnotemark[1]}
\date{\today}

\maketitle
\renewcommand{\thefootnote}{\fnsymbol{footnote}}
\footnotetext[1]{School of Mathematical Sciences, Fudan University,
Shanghai 200433, China and Department of Mathematics, University of
Leuven (KU Leuven), Celestijnenlaan 200B, B-3001 Leuven, Belgium.
E-mail: lun.zhang\symbol{'100}wis.kuleuven.be and
lorencheung@gmail.com This author is a Postdoctoral Fellows of the
Fund for Scientific Research - Flanders (FWO), Belgium.}

\begin{abstract}
The product of $M$ complex random Gaussian matrices of size $N$ has
recently been studied by Akemann, Kieburg and Wei. They showed that,
for fixed $M$ and $N$, the joint probability distribution for the
squared singular values of the product matrix forms a determinantal
point process with a correlation kernel determined by certain
biorthogonal polynomials that can be explicitly constructed. We find
that, in the case $M=2$, the relevant biorthogonal polynomials are
actually special cases of multiple orthogonal polynomials associated
with Macdonald functions (modified Bessel functions of the second
kind) which was first introduced by Van Assche and Yakubovich. With
known results on asymptotic zero distribution of these polynomials
and general theory on multiple orthogonal polynomial ensembles, it
is then easy to obtain an explicit expression for the distribution
of squared singular values for the product of two complex random
Gaussian matrices in the limit of large matrix dimensions.
\end{abstract}


\section{Introduction and statement of the results}
Products of random matrices have presently attracted the most
interest due to their important applications in statistical physics
relating to disordered and chaotic dynamical systems \cite{CGV}, and
other fields beyond physics like MIMO (multiple-input and
multiple-output) networks in telecommunication \cite{SZ,TV}, etc..
The central issue of the study is to find the distributions of the
eigenvalues or singular values for the products in different regimes
and various methods have been applied to perform the spectral
analysis; cf. recent physical literatures
\cite{AB,AkIpKi,AKW,AS,BJLNS,BJW,Ip,PZ} and mathematical literatures
\cite{BBCC,GT}.

This note originates from a recent paper by Akemann, Kieburg and Wei
\cite{AKW}, where they considered the correlation functions for the
products of $M$ quadratic random matrices with complex elements and
no symmetry. More precisely, let $X_j$, $j=1,\ldots,M$ be
independent complex matrices of size $N$ with identical, independent
Gaussian distribution
\begin{equation}\label{eq:Gaussian distribution}
\exp\left[-\textrm{Tr}(X_j^{\ast}X_j)\right],
\end{equation}
where the superscript $^\ast $ stands for conjugate transpose. The
interest lies in the singular values of the product matrix $P_M$
defined by
\begin{equation}\label{def:PM}
P_M:=X_MX_{M-1}\ldots X_1.
\end{equation}
Note that for $M=1$, it is the well-known chiral Gaussian Unitary
Ensemble. By using the change of variable and
Harish-Chandra-Itzykson-Zuber (HCIZ) integral formula, it is shown
that the joint probability density function $\mathcal
{P}_{\textrm{jpdf}}(x_1,\ldots,x_N)$ for the squared singular values
of $P_M$ is given by (see \cite[Equation~(2.13)]{AKW})
\begin{equation}\label{eq:1mm density}
C_N^{(M)}\prod_{1\leq i<j\leq N}(x_j-x_i)\det_{1\leq i, j\leq
N}\left[G_{0,M}^{M,0} \left({-\atop
0,\cdots,0,j-1}\Big{|}x_i\right)\right],
\end{equation}
where
$$\left(C_N^{(M)}\right)^{-1}=N!\prod_{i=1}^M \Gamma (i)^{M+1}$$
is the normalization constant and $G_{0,M}^{M,0}$ is the so-called
Meijer $G$-function which has an integral representation involving
Gamma functions (cf. \cite[Chapter 16]{DLMF}). The formula
\eqref{eq:1mm density} is also called ``one-matrix'' representation
in \cite{AKW}.

Since the matrix inside the determinant in \eqref{eq:1mm density} is
labeled by indices of the Meijer $G$-fucntion, it is not convenient
for the computation of arbitrary $k$-point correlation function. To
overcome this difficulty, they derived an alternative ``two-matrix''
formulation in the sense of \cite{ADOS}, which is the joint
probability density function for the squared singular values of a
single matrix $X_1$ and the squared singular values of the entire
product matrix $P_M$. This setting leads to the other representation
of $\mathcal{P}_{\textrm{jpdf}}(x_1,\ldots,x_N)$ with the help of
the following biorthogonal polynomials.

Let $p_{j}^{(M)}(x)$ and $q_{k}^{(M)}(y)$ be two sequences of monic
polynomials (i.e., with leading coefficient one) of degree $j$ and
$k$ respectively, and satisfy the biorthogonality relations
\begin{equation}\label{biorth}
\int_0^{\infty}\!\!\int_0^{\infty}
w^{(M)}(x,y)p_{j}^{(M)}(x)q_{k}^{(M)}(y)\ud x \ud y =\delta_{j,k}
h_j^{(M)}, \quad j,k=0,1,2,\cdots,
\end{equation}
with the squared norms $h_j^{(M)}=(j!)^{M+1}$ and the weight
function as
\begin{equation}\label{eq:weight}
w^{(M)}(x,y) = y^{-1}e^{-y}G_{0,M-1}^{M-1,0}\left({-\atop
0,\ldots,0}\Big{|}\frac{x}{y}\right), \quad M>1;
\end{equation}
see \cite[Equations~(3.2)~and~(3.3)]{AKW}. In particular, one has
for $M=2$,
\begin{equation}\label{eq:weight M=2}
w^{(2)}(x,y) = y^{-1}e^{-y}G_{0,1}^{1,0}\left({-\atop
0}\Big{|}\frac{x}{y}\right)=y^{-1}e^{-\frac{x}{y}-y}.
\end{equation}
The formulas for  $p_{j}^{(M)}(x)$ and $q_{k}^{(M)}(y)$ can be
derived explicitly. Indeed, we have that
\begin{equation}
q_k^{(M)}(x)=(-1)^kk!L_k(x)
\end{equation}
is the monic Laguerre polynomials associated with weight $e^{-t}$,
where $L_k$ is the standard Laguerre polynomials of degree $k$ (cf.
\cite[Chapter 18]{DLMF}), and
\begin{equation}\label{eq:pj}
p_k^{(M)}(x)=\sum_{j=0}^k\frac{(-1)^{k-j}}{(k-j)!}\left(\frac{k!}{j!}\right)^{M+1}x^j;
\end{equation}
see \cite[Equations~(3.19)~and~(3.21)]{AKW}.

The squared singular values of $P_M$ then form a determinantal point
process of the form
\begin{equation}\label{eq:det point process}
\mathcal{P}_{\textrm{jpdf}}(x_1,\ldots,x_N)=\frac{1}{N!}\det_{1\leq
i,j\leq N}\left[K_N^{(M)}(x_i,x_j)\right],
\end{equation}
where $K_N^{(M)}$ is the correlation kernel given by
\begin{equation}\label{kernel:Kn}
K_N^{(M)}(x,y) = \sum_{j=0}^{N-1} \frac{1}{h_j^{(M)}}
\left(\int_0^{\infty} w^{(M)}(x,s)q_{j}^{(M)}(s) \ud
s\right)p_{j}^{(M)}(y)
\end{equation}
with $h_j^{(M)}$ as in \eqref{biorth}. It is shown in \cite{AKW}
that the two representations \eqref{eq:1mm density} and
\eqref{eq:det point process}--\eqref{kernel:Kn} are equivalent.

A natural question now is to establish the limiting mean
distribution of the squared singular values for $P_M$. It is
observed in \cite{AKW} that, after proper scaling, the macroscopic
limit of the correlation kernel $K_N^{(M)}$ exists for each $M$. For
$M=1$, it is given by the famous Marchenko-Pastur density, as
expected. For $M\geq 2$, however, it is mentioned there the explicit
formula is difficult to derive. Alternatively, they showed the
algebraic equation of degree $M+1$ satisfied by the Stieltjes
transform of the limit, following the approach in \cite{BJLNS}. The
relevant solution of the equation for the case $M=2$ is presented as
well. It is the aim of this note to point out that for $M=2$, we
still have an explicit formula for the limit. Our main theorem is
stated as follows.
\begin{theorem} \label{th: noncrit limit}
Let $X_1$ and $X_2$ be two independent complex matrices of size $N$
with Gaussian distribution as in \eqref{eq:Gaussian distribution}.
If we denote by $\lambda_i$, $i=1,\ldots,N$ the squared singular
values of $X_1X_2$ (i.e., $M=2$ in \eqref{def:PM}), then, almost
surely, the associated scaled empirical measure
\begin{equation}\label{eq:empirical measure}
\mu_N:=\frac{1}{N}\sum_{i=1}^{N}\delta_{\lambda_i/N^2},
\end{equation}
where $\delta_x$ stands for the Dirac point mass at $x$, converges
weakly and in moments to a probability measure $\mu$ on the positive
real axis with density
\begin{equation}\label{def:mu}
\frac{\ud\mu}{\ud x}(x)=\begin{cases}
\tfrac{4}{27}h(\tfrac{4x}{27}), & x\in(0,\tfrac{27}{4}), \\
0, &elsewhere,\end{cases}
\end{equation} where
\begin{equation}\label{def:h}
h(y)=\frac{3\sqrt{3}}{4\pi}
\frac{(1+\sqrt{1-y})^{1/3}-(1-\sqrt{1-y})^{1/3}}{y^{2/3}},\quad
0<y<1,
\end{equation}
as $N \to \infty$.
\end{theorem}
This theorem also implies the statement
\[
\lim_{N \to \infty}NK_N^{(2)}(N^2x,N^2x)=\frac{\mathrm d
\mu}{\mathrm d x}(x), \qquad x>0,
\]
where $K_N^{(2)}(x,y)$ is defined in \eqref{kernel:Kn}. We point out
that the density \eqref{def:mu}--\eqref{def:h} has been found
independently in several other unrelated problems (see
\cite{LSZ,PS}), and therefore seems to enjoy a certain universality.

The rest of this note is mainly devoted to the proof of Theorem
\ref{th: noncrit limit}. The essential issue is the observation that
the biorthogonal polynomials $p_k^{(2)}$ in \eqref{eq:pj} are
actually certain \emph{multiple orthogonal polynomials} (cf.
Nikishin and Sorokin \cite[Chapter 4, \S 3]{NikSor}, Ismail
\cite[Chapter 23]{Ismail}) that have already been studied by the
community of orthogonal polynomials. In view of the natural
appearances of multiple orthogonal polynomials in certain random
models from mathematical physics over the past few years, including
random matrix theory, non-intersecting paths, etc. (cf.
\cite{Kuijl,Kuijl2} and references therein), this note provides
another example that bridges two different groups. We end this note
with some concluding remarks.

\section{Proof of the main theorem}

\subsection{Multiple orthogonal polynomials associated with Macdonald functions}
Multiple orthogonal polynomials are polynomials of one variable
which are defined by orthogonality relations with respect to $r$
different measures, where $r \geq 1$ is a positive integer. They can
be viewed as a generalization of orthogonal polynomials, which have
important applications in approximation theory and number theory;
cf. \cite{Apt,NikSor,WVA99}.

For our purpose, let us define a function $\rho_\gamma$ by
\begin{equation}
\rho_\gamma(x)=2x^{\gamma/2}K_\gamma(2\sqrt{x}), \qquad x>0,
\end{equation}
where $K_\gamma$ is the Macdonald function (modified Bessel function
of the second kind; see \cite[Chapter 10]{DLMF}) and consider two
weights
\begin{equation}\label{two measures}
d\mu_1(x)=x^{\kappa}\rho_\gamma(x) \ud x,\quad
d\mu_2(x)=x^{\kappa}\rho_{\gamma+1}(x) \ud x,\quad \kappa>-1, \quad
\gamma\geq 0,
\end{equation}
on the positive real line. For any vector $(k,m)\in\mathbb{N}^2$,
the type II multiple orthogonal polynomials
$p_{k,m}^{(\gamma,\kappa)}$ for the system of weights
$(\mu_1,\mu_2)$ are such that $p_{k,m}^{(\gamma,\kappa)}$ is a monic
polynomial of degree $k+m$ and satisfies the following multiple
orthogonality conditions:
\begin{align}
\int_0^{\infty}p_{k,m}^{(\gamma,\kappa)}(x)x^j d\mu_1(x)&=0, \quad
j=0,1,...,k-1,
\\
\int_0^{\infty}p_{k,m}^{(\gamma,\kappa)}(x)x^j d\mu_2(x)&=0, \quad
j=0,1,...,m-1.
\end{align}
By taking $m=k$, we set
\begin{align*}
P_{2k}^{(\gamma,\kappa)}(x)=p_{k,k}^{(\gamma,\kappa)}(x),\qquad
P_{2k+1}^{(\gamma,\kappa)}(x)=p_{k+1,k}^{(\gamma,\kappa)}(x).
\end{align*}
An explicit formula for $P_k^{(\gamma,\kappa)}$ is given by
\begin{equation}\label{Pn}
P_k^{(\gamma,\kappa)}(x)=\sum_{j=0}^{k}a_k(j)x^{k-j},
\end{equation}
where
\begin{equation}\label{eq:ak}
a_k(j)=(-1)^j\binom{k}{j}\frac{(\kappa+1)_k(\kappa+\gamma+1)_k}
{(\kappa+1)_{k-j}(\kappa+\gamma+1)_{k-j}},\quad 0\leq j \leq k;
\end{equation}
see \cite[Theorem 2]{CVA2001}. The measures $(d\mu_1,d\mu_2)$ from
\eqref{two measures} form an AT system (cf. \cite{NikSor}), that is,
for any index $(k,m)\in\mathbb{N}^2$, every linear combination of
the functions
\[ \{
w_1,xw_1,\ldots,x^{k-1}w_1,w_2,xw_2,\ldots,x^{m-1}w_2\} \] has at
most $k+m-1$ zeros on the positive real axis, where
$w_1(x)=x^{\kappa}\rho_\gamma(x)$ and
$w_2(x)=x^{\kappa}\rho_{\gamma+1}(x)$. As a consequence, all the
zeros of $P_k^{(\gamma,\kappa)}$ are simple, lie in $(0,+\infty)$,
and satisfy the interlacing property \cite{CCVA}. Furthermore, it
was shown in \cite{VAY} that $P_k^{(\gamma,\kappa)}$ satisfies the
following four-term recurrence relation
\begin{equation}
\label{eq:recurrence_relation} x P_k^{(\gamma,\kappa)}(x) =
P_{k+1}^{(\gamma,\kappa)}(x) + b_k P_k^{(\gamma,\kappa)}(x) + c_k
P_{k-1}^{(\gamma,\kappa)}(x) + d_k P_{k-2}^{(\gamma,\kappa)}(x)
\end{equation}
with recurrence coefficients
\begin{equation} \label{eq:coefficientsbcdk}
\begin{aligned}
    b_{k} & = (k+\kappa+1)(3k+\kappa+2\gamma) - (\kappa+1)(\gamma-1), \\
    c_{k} & = k (k+\kappa) (k+\kappa+\gamma) (3k+2\kappa+\gamma), \\
    d_{k} & = k(k-1)(k+\kappa-1)(k+\kappa)(k+\kappa+\gamma-1)(k+\kappa+\gamma).
    \end{aligned}
\end{equation}

These polynomials constitute one of few examples of multiple
orthogonal polynomials that are not related to the classical
orthogonal polynomials. They are first introduced by Van Assche and
Yakubovich in \cite{VAY}, which solve an open problem posed by
Prudnikov \cite{PRU}.

We now introduce a new parameter $n\in \mathbb{N}$ and put
\begin{equation}\label{scaled P(k,n)}
P_{k,n}^{(\gamma,\kappa)}(x):=\frac{P_k^{(\gamma,\kappa)}(n^2
x)}{n^{2k}}.
\end{equation}
Clearly, $P_{k,n}^{(\gamma,\kappa)}(x)$ is a monic polynomial of
degree $k$ for each $n$. For each $P_{k,n}^{(\gamma,\kappa)}$, we
can associate the normalized counting zero measure defined by
\begin{equation}\label{eq:counting measure}
\nu(P_{k,n}^{(\gamma,\kappa)})=\frac{1}{k}\sum_{P_{k,n}^{(\gamma,\kappa)}(x)=0}
\delta_{x}.
\end{equation}
A measure $\nu^{\xi}$ is called the asymptotic zero distribution of
$\{P_{k,n}^{(\gamma,\kappa)}\}$ if
\begin{equation} \label{eq:weak conv}
\lim_{k/n \to \xi} \int
f \, \ud\nu(P_{k,n}^{(\gamma,\kappa)})=\int f \, \ud\nu^{\xi}
\end{equation} for every bounded continuous function $f$ on $\mathbb{R}$,
i.e., it is the weak limit of the measures
$\nu(P_{k,n}^{(\gamma,\kappa)})$. Here the notation $\lim_{k/n \to
\xi}$ means that both $k,n\to \infty$ with $k/n\to \xi>0$.

The explicit formula for $\nu^{\xi}$ is established by Coussement,
Coussement and Van Assche as stated in the following proposition
(see \cite[Theorem 2.7]{CCVA}).
\begin{proposition}\label{prop:zero distribution}
The asymptotic zero distribution of the multiple orthogonal
polynomials associated with Macdonald functions \eqref{scaled
P(k,n)} exists and has the density
\begin{equation}
\frac{\ud\nu^{\xi}}{\ud x}(x)=\begin{cases}
\tfrac{4}{27\xi^2}h(\tfrac{4x}{27\xi^2}), & x\in(0,\tfrac{27\xi^2}{4}), \\
0, &elsewhere,\end{cases}
\end{equation} where
\begin{equation}
h(y)=\frac{3\sqrt{3}}{4\pi}
\frac{(1+\sqrt{1-y})^{1/3}-(1-\sqrt{1-y})^{1/3}}{y^{2/3}}, \quad
0<y<1.
\end{equation}
\end{proposition}

\begin{remark}
In practice, one can also scale the parameters $(\kappa,\gamma)$,
say, putting $\kappa \mapsto pn$ and $ \gamma \mapsto qn$ with
$p,q>0$. The case when $\kappa$ and $\gamma$ are fixed corresponds
to taking the limits $p,q \to 0$, respectively. Under this general
setting, the normalized zero counting measure \eqref{eq:counting
measure} converges weakly to the first component of a vector of two
measures which satisfies a vector equilibrium problem with two
external fields and the explicit formula for the equilibrium vector
is given in terms of solutions of an algebraic equation; see
\cite{ZP} for details. The vector equilibrium problem that
characterizes $\nu^{1}$ is also presented in Section 3 below.
\end{remark}

With the above preparations, we are ready to prove Theorem \ref{th:
noncrit limit}.

\subsection{Proof of Theorem \ref{th: noncrit limit}}
A comparison of \eqref{eq:pj} and \eqref{Pn}--\eqref{eq:ak}
immediately gives that
\begin{equation}\label{eq:bi as mul}
p_{k}^{(2)}(x)=P_{k}^{(0,0)}(x),
\end{equation}
thus, $p_{k}^{(2)}$ are multiple orthogonal polynomials with respect
to two weight functions $(2K_0(2\sqrt{x}),2K_1(2\sqrt{x}))$ for
$x>0$. This, together with \eqref{eq:counting measure} and
Proposition \ref{prop:zero distribution} implies that, as $N\to
\infty$, the measure
\begin{equation}\label{def:nuN}
\nu_N:=\frac{1}{N}\sum_{p_N^{(2)}(x)=0}\delta_{x/N^2}=\nu(P_{N,N}^{(0,0)})
\end{equation}
converges weakly to $\nu^{1}$, which agrees with the measure $\mu$
defined in Theorem \ref{th: noncrit limit}.

To show the measure $\mu$ is the almost sure weak convergence of the
empirical measure \eqref{eq:empirical measure}, we note that
$p_{N}^{(2)}$ is actually the average characteristic polynomials
with respect to the multiple orthogonal polynomial ensemble
\eqref{eq:det point process}, i.e.,
$$
p_{N}^{(2)}(z)=\mathbb{E}\left[\prod_{i=1}^N(z-x_i)\right], \qquad
z\in\mathbb{C},
$$
where $\mathbb{E}$ refers to \eqref{eq:det point process} with
$M=2$; cf. \cite[Proposition 2.2]{Kuijl}. For a determinantal point
process on $\mathbb{R}$ that belongs to biorthogonal ensembles
\cite{Bor}, a sufficient condition that the limiting zero
distribution of the average characteristic polynomials coincides
with the almost sure weak convergence of the empirical measure of
this determinantal point process is recently established by Hardy
\cite{Hardy}, which in particular includes multiple orthogonal
polynomial ensembles as special cases. By \cite[Theorem 1.3 and
Corollary 1.5]{Hardy}, it suffices to show that
\begin{itemize}
  \item the recurrence coefficients for
$p_N^{(2)}(n^2x)/n^{2N}=P_{N,n}^{(0,0)}(x)$ are bounded for every
$N,n$ and certain $\varepsilon>0$ such that $|\frac{N}{n}-1|\leq
\varepsilon$;
  \item $\nu(P_{N,n}^{(0,0)})$ defined in \eqref{eq:counting measure} converges to $\nu^{\xi}$ in
  moments, i.e., for any $l\in\mathbb{N}$,
  \begin{equation}
  \lim_{N/n \to \xi} \int
x^l \, \ud\nu(P_{N,n}^{(0,0)})=\int x^l \, \ud\nu^{\xi},
  \end{equation}
  where $\xi\in[0,1+\varepsilon)$.
\end{itemize}
These two conditions are easily verified in our case. Indeed, from
\eqref{eq:recurrence_relation} and \eqref{eq:coefficientsbcdk}, one
has
\begin{multline}
\label{eq:recurrence_P_doubly_indexed} x P_{N,n}^{(0,0)}(x) =
P_{N+1,n}^{(0,0)}(x) + b_{N,n}P_{N,n}^{(0,0)}(x)\\ + c_{N,n}
P_{N-1,n}^{(0,0)}(x) + d_{N,n} P_{N-2,n}^{(0,0)}(x),
\end{multline}
with recurrence coefficients given by
\begin{equation} \label{eq:coefficientsbcdkn}
\begin{aligned}
    b_{N,n} & = \frac{3N(N+1)-1}{n^2}\to 3\xi^2, \\
    c_{N,n} & = \frac{3N^4}{n^4}\to 3\xi^4, \\
    d_{N,n} & = \frac{N^3(N-1)^3}{n^6} \to \xi^6,
    \end{aligned}
    \end{equation}
if $N/n\to\xi$. Thus, there exists $\varepsilon>0$, such that
\begin{equation}\label{eq:bounds of recurr}
\max _{|\frac{N}{n}-1|\leq
\varepsilon}\left(|b_{N,n}|,|c_{N,n}|,|d_{N,n}|\right)<+\infty.
\end{equation}
The equations \eqref{eq:recurrence_P_doubly_indexed} and
\eqref{eq:coefficientsbcdkn} also implies that
$P_{N,n}^{(0,0)}(\lambda)=0$ if and only if $\lambda$ is an
eigenvalue of the banded Toeplitz matrix $T_N$ defined by
\begin{equation}
T_N=\begin{pmatrix} b_{N,n} & 1 & 0 & \ldots & \ldots & 0 \\
c_{N,n} & b_{N,n} & 1 & \ddots & & \vdots \\
d_{N,n} & c_{N,n} & b_{N,n} & 1 & \ddots & \vdots \\
0 & d_{N,n} & c_{N,n} & b_{N,n} & \ddots & 0 \\
\vdots & \ddots & \ddots & \ddots & \ddots & 1 \\
0 & \ldots & 0 & d_{N,n} & c_{N,n} & b_{N,n}
\end{pmatrix}_{N\times N}.
\end{equation}
By Gershgorin circle theorem, it follows that the spectral radius
$\rho_N$ of $T_N$ is bounded by $$\sup_{N\leq
n}|b_{N,n}|+\sup_{N\leq n}|c_{N,n}|+\sup_{N\leq n}|d_{N,n}|+1.$$
This, together with \eqref{eq:bounds of recurr}, implies
$\sup_N\rho_N<+\infty$. We then have that the support of
$\nu(P_{N,n}^{(0,0)})$ is uniformly bounded. Since it is already
known that $\nu(P_{N,n}^{(0,0)})$ converges weakly to $\nu^{\xi}$,
we obtain the convergence of $\nu(P_{N,n}^{(0,0)})$ to $\nu^{\xi}$
in moments.

This completes the proof of our main theorem.

\section{Concluding remarks}
In this note, we have given the explicit formula for the limiting
distribution of squared singular values for the products of two
complex random Gaussian matrices. Our approach is based on the
finite-$N$ density derived in \cite{AKW} and identifying the
relevant biorthogonal polynomials with the special cases of multiple
orthogonal polynomials associated with Macdonald functions. It would
be interesting to see whether some other multiple orthogonal
polynomials will appear for the products of $M>2$ complex Gaussian
matrices; see also \cite{AkIpKi} for a study of products of
rectangular random matrices.

The next challenge is to establish the microscopic limit of the
correlation kernel \eqref{kernel:Kn}. From Theorem \ref{th: noncrit
limit}, it is readily seen that, for $M=2$, the density function of
the limiting mean distribution vanishes like a squared root at the
soft edge $27/4$, and blows up of order $x^{-2/3}$ at the hard edge.
This observation leads to the conjecture to, as posed in \cite{AKW},
the universal sin and Airy kernels for the bulk and soft edge of the
spectrum, but new universality classes are required for the
description of the local behavior at the hard edge. Due to the
appearance of multiple orthogonal polynomials, a possible way to
tackle this problem is to perform Deift/Zhou steepest descent
analysis \cite{Dei} for the associated Riemann-Hilbert problem
\cite{VAGK}. To this end, it is worthwhile to mention an alternative
characterization of the measure $\mu$ given in Theorem \ref{th:
noncrit limit}. Consider the energy functional defined by
\begin{align*} \label{functionalwithV}
    \iint &\log \frac{1}{|x-y|} d\nu_1(x)d\nu_1(y) +
    \iint \log \frac{1}{|x-y|}d\nu_2(x)d\nu_2(y) \nonumber \\
    &-\iint \log \frac{1}{|x-y|} d\nu_1(x)d\nu_2(y)
    + \int \sqrt{4x} d\nu_1(x).
\end{align*}
We want to minimize this functional over all vectors of measures
$(\nu_1,\nu_2)$ such that $\supp(\nu_1)\subset [0,\infty)$, $\int
d\nu_1 = 1$ and $\supp(\nu_2)\subset (-\infty,0]$, $\int d\nu_2 =
1/2$. It was shown by Rom\'{a}n and the author in \cite{ZP} that the
measure $\mu$ is the first component of the unique minimizer. We
believe such kind of characterization will play a important role in
the nonlinear steepest descent analysis of the Riemann-Hilbert
problem.

\section*{Acknowledgements}
I would like to thank Adrien Hardy and Arno Kuijlaars for helpful
discussions, and the referee for careful reading and constructive
suggestions.

\end{document}